\documentclass[runningheads]{llncs}
\usepackage[dvipdfmx]{graphicx}
\usepackage{amsmath, amssymb}

\usepackage{algorithm,algpseudocode}
\usepackage{etoolbox}
\makeatletter

\newcommand*{\algrule}[1][\algorithmicindent]{%
  \makebox[#1][l]{%
    \hspace*{.2em}
    \vrule height .75\baselineskip depth .25\baselineskip
  }
}

\newcount\ALG@printindent@tempcnta
\def\ALG@printindent{%
    \ifnum \theALG@nested>0
    \ifx\ALG@text\ALG@x@notext
    \else
    \unskip
    \ALG@printindent@tempcnta=1
    \loop
    \algrule[\csname ALG@ind@\the\ALG@printindent@tempcnta\endcsname]%
    \advance \ALG@printindent@tempcnta 1
    \ifnum \ALG@printindent@tempcnta<\numexpr\theALG@nested+1\relax
    \repeat
    \fi
    \fi
}
\patchcmd{\ALG@doentity}{\noindent\hskip\ALG@tlm}{\ALG@printindent}{}{\errmessage{failed to patch}}
\patchcmd{\ALG@doentity}{\item[]\nointerlineskip}{}{}{} 
\makeatother

\algnewcommand\algorithmicinput{\textbf{Input:}}
\algnewcommand\algorithmicoutput{\textbf{Output:}}
\algnewcommand\Input{\item[\algorithmicinput]}%
\algnewcommand\Output{\item[\algorithmicoutput]}%

\begin{document}

\title{Improved Scheduling of Morphing Edge Drawing}
\titlerunning{Improved Scheduling of Morphing Edge Drawing}
\author{Kazuo Misue\orcidID{0000-0003-0216-8969}}
\authorrunning{K. Misue}
\institute{University of Tsukuba, Tsukuba, Japan\\
\email{misue@cs.tsukuba.ac.jp}}

\maketitle

\begin{abstract}
Morphing edge drawing (MED), a graph drawing technique, is a dynamic extension of partial edge drawing (PED), where partially drawn edges (stubs) are repeatedly stretched and shrunk by morphing.
Previous experimental evaluations have shown that the reading time with MED may be shorter than that with PED.
The morphing scheduling method limits visual clutter by avoiding crossings between stubs.
However, as the number of intersections increases, the overall morphing cycle tends to lengthen in this method, which is likely to have a negative effect on the reading time.
In this paper, improved scheduling methods are presented to address this issue. 
The first method shortens the duration of a single cycle by overlapping a part of the current cycle with the succeeding one.
The second method duplicates every morph by the allowable number of times in one cycle.
The third method permits a specific number of simultaneous crossings per edge.
The effective performances of these methods are demonstrated through experimental evaluations. 

\end{abstract}

\section{Introduction}

Partial edge drawing (PED) is a graph-drawing technique in which the edges are drawn partially to avoid crossings.
Morphing edge drawing (MED) is a dynamic graph representation technique in which the stubs (partially drawn edges) are repeatedly stretched and shrunk by morphing \cite{misue2019-graph}.
Experiments by Bruckdorfer have suggested that, compared with graph drawings in which edges are drawn as complete line segments, PED may improve the reading accuracy and increase the reading time \cite{bruckdorfer2012mad}.
An experimental evaluation by Misue \& Akasaka showed that MED has the potential to reduce reading time compared to PED \cite{misue2019-graph}.
It is possible that, as the stubs change with morphing, less time is needed to guess the erased parts.
The scheduling method shown by Misue \& Akasaka schedules MED morphing so that stubs do not create new crossings.
In other words, in a situation where two edges intersect, while one stub is stretched, the other must wait as short.
Although this type of scheduling maintains the reduction of visual clutter by PED, it forces the morphing cycle to increase as the number of nodes, edges, and intersection points increases.
Here, the morphing cycle is the total of the morphing time of all edges.
Correspondingly, the latency before morphing used to determine whether two nodes are adjacent to each other may be longer than the time needed for guessing.

To address this issue, three methods to shorten the morphing cycle in MED were developed in this study, as explained below:
The first method shortens the duration of a single cycle by initiating the new morphing of some stubs without waiting for all the stubs to regain their shortest states.
In fact, in the MED scheduling shown by Misue \& Akasaka, the duration of one cycle starts when all the stubs are in their shortest states and eventually ends when they all return to their initial shortest states again.
Here, we have exploited the idea that, even if the morphing of some stubs begins before all the stubs return to their shortest states, no crossing may occur, and the duration of a cycle is reduced.
In the second method, every morph is duplicated by the allowable number of times in one cycle.
In previous MED scheduling, each edge stub is stretched and shrunk only once within each cycle.
However, some stubs can be morphed two or more times within one cycle without leading to crossing.
Considering this, multiple morphings within a single cycle can shorten the average duration of a morphing cycle.
Finally, in the third method, crossings between edges are allowed to occur. Although graph drawings with many crossings are difficult to read, a small number of crossings are considered to have only a limited impact on readability~\cite{purchase1996validating,purchase1997which}.
Therefore, we developed a scheduling method that allows up to a certain number of simultaneous crossings per edge.

The contributions of this study can be summarized as follows:

\begin{enumerate}
\item Three new scheduling methods were presented to shorten the morphing cycle in MED, and
\item The effectiveness of each scheduling method is demonstrated experimentally.
\end{enumerate}

\section{Partial Edge Drawing and Morphing Edge Drawing}

A simple undirected graph is denoted by $G=(V,E)$ and the drawing of a graph $G$ is denoted by $\Gamma(G)=(\Gamma(V),\Gamma(E))$.
$\Gamma(V)=\{\Gamma(v)|v\in V\}$ and $\Gamma(E)=\{\Gamma(e)|e\in E\}$.
Herein, $\Gamma(G)$ is a traditional straight-line drawing, $\Gamma(v)$ of node $v\in V$ is a point located at position $p_{v}$, and $\Gamma(e)$ of edge $e\in E$ is a line segment connecting two nodes (points).
In other words, it can be expressed as $\Gamma(e)=\{s\cdot p_{w}+(1-s)\cdot p_{v}|s\in [0,1]\}$, where $e=\{v,w\}$.
Drawing $\Gamma(G)$ can be referred to by the retronym {\em complete edge drawing (CED)} because it completely draws a straight-line segment to represent an edge.
The layout of graph $G$, that is, $\Gamma(G)$, is assumed to be provided in advance within this study.
To simplify the description in subsequent sections, $\Gamma$ is omitted and $e$ is used to replace $\Gamma(e)$ when it is clear from the context that it represents $\Gamma(e)$.

\subsection{Partial Edge Drawing}

The partial drawing of the edge $e=\{v,w\}$ is represented by the function $\gamma_{e}:[0,1]^2 \rightarrow 2^{\Gamma(e)}$, as shown in Eq.~(\ref{eq:gammae}).
\begin{equation}
  \gamma_{e}(\alpha,\beta) =
  \begin{cases}
    \{s\cdot p_{w}+(1-s)\cdot p_{v}|s\in [0,\alpha]\cup [\beta,1]\} & \text{for $\alpha < \beta$} \\    
    \Gamma(e)    & \text{for $\alpha \ge \beta$}.
  \end{cases}  
\label{eq:gammae}
\end{equation}
The partial drawing $\gamma_{e}(\alpha,\beta)$ of edge $e$ is the remainder of the entire $\Gamma(e)$ mapped to the interval $[0, 1]$, with the part corresponding to interval $(\alpha, \beta)$ removed from $\Gamma(e)$.
Each of the remaining contiguous parts is called a {\em stub}.
If the part to be deleted is not the end of edge $\Gamma(e)$, that is, if $0<\alpha$ and $\beta<1$, two stubs remain at the two nodes incident to the edge $e$.
We refer to them as a {\em pair of stubs}.

Given $\alpha_{e}$ and $\beta_{e}$ for all edges $e\in E$ and that there exists an edge $e_{1}\in E$ such that $\alpha_{e_{1}}<\beta_{e_{1}}$, the drawing $\Gamma_{PED}(G)=(\Gamma(V),\Gamma_{PED}(E))$ is called a {\em PED}, where $\Gamma_{PED}(E)=\{\gamma_{e}(\alpha_{e},\beta_{e})|e\in E\}$.
When the lengths of a pair of stubs are equal, that is, when there exists a relationship $\alpha_{e} = 1-\beta_{e}$, this is called a {\em symmetric PED (SPED)}.
In this case, the smaller parameter $\alpha_{e}$ is called the {\em stub ratio}.
If the stub ratios for all edges are the same $\delta$, the drawing is called {\em $\delta$-symmetric homogeneous PED ($\delta$-SHPED)}.

\subsection{Morphing Edge Drawing}

Let $T$ be a set of times.
A dynamic drawing $\Gamma_{MED}(G)=(\Gamma(V),\Gamma_{MED}(E))$ , which is constructed using the morphing function $\mu_{e}:T \rightarrow 2^{\Gamma(e)}$ and defines the partial drawing of edge $e$ at time $t\in T$.
It is called the {\em morphing edge drawing (MED)}, where $\Gamma_{MED}(E)=\{\mu_{e}|e\in E\}$.
Let $\rho_{e}: T\rightarrow [0,1]^{2}$ be a function that defines the parameters of the partial edge for time $t\in T$. The function $\mu_{e}$ can then be constructed as $\mu_{e}(t) = \gamma_{e}(\rho_{e}(t))$.
For all edges $e\in E$, if the function $\rho_{e}$ satisfies $\rho_{e}(t)=(\delta_{t}, 1-\delta_{t})$ (where $0\le \delta_{t} \le 1/2$) for $\forall t\in T$, then a SPED is obtained at any time.
MED constructed based on such function is called a {\em symmetric MED (SMED)}.

This study focuses on SMED.
In one morph, each stub changes from the shortest state (stub-ratio $\delta$) to the longest state (stub-ratio $\eta$), remains in the longest state for a certain time and then returns to the shortest state.
The range over which a stub stretches and shrinks is called as the {\em morphing range}.
Two paired stubs start and end morphing simultaneously.

\section{Related Work}

Bruckdorfer et al. have given the formulation of PED \cite{bruckdorfer2012mad}.
Burch et al. \cite{burch2012evaluating} demonstrated the applicability of this approach to directed graphs using tapered links to represent partially drawn edges.
Schmauder et al. applied PED to weighted graphs by coloring edges to represent their weights \cite{schmauder2015visualizing}.
Information on PED is summarized in the commentary by N\"{o}llenburg \cite{nollenburg2020-crossing}.

Bruckdorfer et al. conducted experiments comparing CED and 1/4-SHPED with respect to graph-reading performance \cite{bruckdorfer2015ped}.
Although not statistically significant, the chart visualizing the results of the experiment indicated a slightly more accurate but longer response time for 1/4-SHPED than for CED in terms of the graph reading task.
Binucci et al. \cite{binucci2016partial} conducted more detailed evaluation experiments and found that, among the SPEDs, SHPED yielded the best reading accuracy. 
Burch \cite{burch2017user} examined the effects of stub orientation and length on the graph reading accuracy and found that shorter stub lengths tend to result in more misjudgments regarding the target nodes and that stub orientation also affects accuracy.

MED was proposed by Misue \& Akasaka \cite{misue2019-graph}.
The formalization of the MED is provided herein, and evaluation experiments on the readability of the MED indicate that the MED may be superior to the PED, in terms of the reading time.
The formalization presented in Section 2 is based on the one proposed by Misue \& Akasaka \cite{misue2019-graph}.

\section{Terminology and Notation}

This section describes the terminology and notations used in this paper.

\subsection{Set and Set Family}
\label{ssec:term_set}

The sets and functions that return a set are capitalized.
Let $\#(A)$ denote the number of elements in a finite set $A$, $A^{c}$ denote the complement of set $A$, and $2^{A}$ denote the power set of set $A$.
For set $A$, let $A^{\#k}$ denote the set family created by collecting only all the subsets with $k$ ($\geq 1$) elements.
In other words, $A^{\#k}= \{A'\in 2^{A}|\#(A')=k\}$.

\subsection{Time Periods}
\label{ssec:timing}

Suppose that the time period is a subset of $U=(-\infty,\infty)$ and can be expressed as $P=\bigcup_{i=1}^{\chi} [a_{i},b_{i})$ ($\chi\geq 0$).
In this case, $b_{i}<a_{j}$ if $i<j$.
In other words, we assumed that the time period can be represented as a union set of noncontiguous half-open intervals.
When $\chi =0$, it is assumed to be empty.

\subsection{Intersections and Types of Intersections}
\label{ssec:intersection}

It can be assumed that the intersection point of two drawn edges can be represented by a pair of edges because the layout of the graph is assumed to be provided in advance.
Therefore, when any two sets of edges cross at the same point, although only one point exists from the geometric perspective, the point is considered as two different intersection points corresponding to the two edge crossings.
If edge $e_{1}$ crosses another edge $e_{2}$ at a point $p$, then, $e_{2}$ is called the opposite edge of $e_{1}$ at $p$ and is denoted by $e_{1}/p$.
In other words, when $e_{1}$ and $e_{2}$ cross at point $p$, $e_{1}/p = e_{2}$ and $e_{2}/p = e_{1}$.
Let $I$ denote an entire set of intersection points.
Furthermore, let $I(e)$ denote a set of intersection points on edge $e$.

\begin{figure}[htb]
\begin{minipage}[t]{0.30\hsize}
\centering
  \includegraphics[scale=0.85]{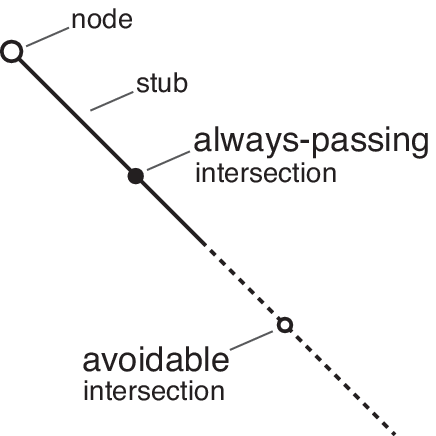}\\
  (a) As seen from one edge
\end{minipage}
\hspace{0.05\hsize}
\begin{minipage}[t]{0.65\hsize}
\centering
  \includegraphics[scale=0.85]{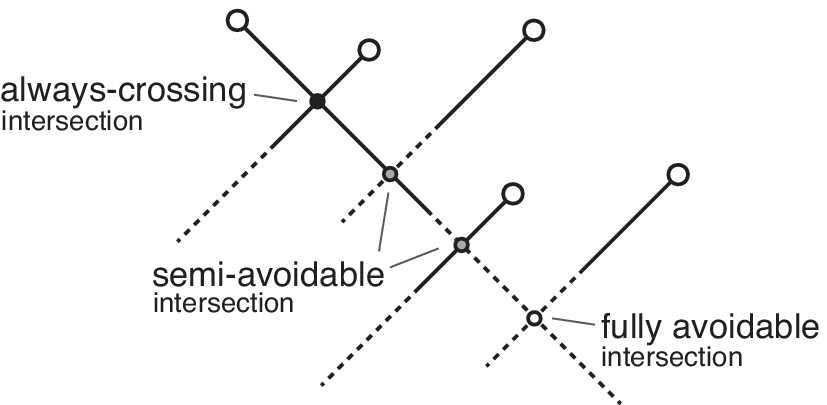}\\
  (b) As seen from two edges
\end{minipage}
  \caption{Type of intersection points. The solid lines represent the state where the stubs are shortest, and the dashed lines represent the state where the stubs are stretched.}
  \label{fig:intersection_types}
\end{figure}

In MED, crossings between stubs may be unavoidable, as illustrated in Fig.~\ref{fig:intersection_types}.
We refer to an intersection point as ``$e$ is {\em always passing}'', where a stub of edge $e$ is passing even when the stub is at its shortest state. 
The intersection points at which both edges always pass are called {\em always crossing}.
The intersection points at which both edges are not always-passing are called {\em fully avoidable}, whereas intersection points at which only one edge is always passing are called {\em semi-avoidable}.
Fully avoidable and semi-avoidable intersections are collectively called {\em avoidable}.
If semi-avoidable intersections are in the morphing range of a stub, the morphing of the stub will always result in one or more crossings.

\section{Scheduling}
\label{sec:scheduling}

The scheduling of a MED defines a morphing function $\mu_{e}:T\rightarrow 2^{\Gamma(e)}$ for all edges $e$.
We assumed that the change in each stub from stretching to shrinking back to the original state with respect to the elapsed time from the start of morphing is already defined by the function $\mu_{e}^{*}:\mathbb{R} \rightarrow 2^{\Gamma(e)}$.
Thus, scheduling implies the determination of the morphing start time for each edge. Once the start time $t_{start}(e)$ has been determined, the morphing function can be defined as $\mu_{e}(t)= \mu_{e}^{*}(t-t_{start}(e))$.

Given a function $\mu_{e}^{*}$, we can determine the elapsed time after the start of morphing to each stub state.
Let $\tau_{pass}(e,p)$ denote the time it takes for the tip of a stub of edge $e$ to pass the intersection point $p$ for the first time (passing while stretching) after the onset of morphing. Let $\tau_{ret}(e, p)$ denote the time it takes for the tip of the stub of edge $e$ to pass the intersection point $p$ for the second time (passing while shrinking) after the onset of morphing.
Let $\tau_{trip}(e)$ denote the time from the start to the end of the morphing of edge $e$.
$\tau_{ret}(e, p)$ and $\tau_{trip}(e)$ include the time when the stub is fully stretched and the morphing is paused.
Let $C^{e}_{p}$ ($\subseteq U$) denote the time period when the stub of edge $e$ passes through point $p$ on $e$.
If the morphing start time of edge $e$ is $t_{start}(e)$, then $C^{e}_{p}=[t_{start}(e) + \tau_{pass}(e, p), t_{start}(e) + \tau_{ret}(e, p))$.
Let $C^{e}_{p}=\emptyset$ if the morphing start time at edge $e$ is undefined.

In the following sections, we first describe the algorithm proposed by Misue \& Akasaka \cite{misue2019-graph} and then extend it in a step-by-step manner to accommodate the overlapping of each cycle, duplicating morphs in one cycle, and allowance of crossings.
In this manner, we proceed with the explanation, while extending the functions.
Thus, we use the numbered function names like $F^{(1)}$ and $F^{(2)}$.
Because the functions with larger numbers are extensions of the smaller-numbered functions, only one function with the largest number needs to be defined for implementation.

\subsection{Basic Scheduling Algorithm}
\label{ssec:basic_scheduling}

Here, all intersection points are assumed to be fully avoidable.

Alg.~\ref{alg:scheduleBasic} shows the algorithm proposed by Misue \& Akasaka \cite{misue2019-graph}. Given a set of edges $E$, this algorithm determines the start time $t_{start}(e)$ of morphing for all edges $e\in E$.
Let $E$ be a morphing group consisting of edges whose morphing timings may affect each other.
The algorithm sequentially determines the start time of morphing with respect to the edges in set $E$.
Misue \& Akasaka \cite{misue2019-graph} sorted the edges in descending order of their lengths and determined the start time of the morphing of each edge in the order.
The method examines, for an edge $e$, the morphing timing of all opposite edges that intersect edge $e$ and have already determined their start time.
It then determines the earliest time at which no crossings occur for edge $e$ as the morphing start time for $e$.
Note that the first morphing is assumed to start at time zero ($t=0$).

\begin{algorithm}[htb]
  \caption{Scheduling morphing}
  \label{alg:scheduleBasic}
  \begin{algorithmic}[1]
    \Input{$E$ -- Set of edges in a morphing group}
    \Output{The start time $t_{start}(e)$ of all $e\in E$, and the total morphing time $t_{total}$}
    \Function{scheduleParallel}{$E$}
      \For{$e$ in $sort(E)$}      
        \State $t_{start}(e) \leftarrow t_{earliest}(P_{fbd}(e))$
      \EndFor
      \State $t_{total} \leftarrow \max_{e\in E} (t_{start}(e) + \tau_{trip}(e))$
    \EndFunction
  \end{algorithmic}
\end{algorithm}

Function $P_{fbd}^{(1)}:E\rightarrow 2^{U}$ provides the time period during which one or more crossings occur when edge $e\in E$ starts morphing.
In other words, $P_{fbd}^{(1)}(e)$ is the forbidden morphing start period used by edge $e$ to avoid crossing with its opposite edges.
The definition of function $P_{fbd}^{(1)}$ is expressed in Eqs.~(\ref{ep:Pfbd1}) and (\ref{ep:Pcrit1}).

\begin{align}
P_{fbd}^{(1)}(e) &= P_{crit}^{(1)}(e) \label{ep:Pfbd1} \\
P_{crit}^{(1)}(e) &= \bigcup_{p\in I_{s}(e)} S^{e}_{p}(C^{e/p}_{p}) \label{ep:Pcrit1}
\end{align}
Let $I_{s}(e)= \{p\in I(e)|t_{start}(e/p)\mbox{ has been defined}\}$.
Function $S^{e}_{p}:2^{U}\rightarrow 2^{U}$ modifies a given time period by the time needed for the stub of edge $e$ to reach the intersection point $p$, extend further, and then return to that point.
$S^{e}_{p}(\emptyset)=\emptyset$, and if $[a,b)\subseteq P$, then $[a-\tau_{ret}(e,p),b-\tau_{pass}(e,p))\subseteq S^{e}_{p}(P)$.
$S^{e}_{p}(P)$ yields the morphing start period for edge $e$ to pass through point $p$ within the time period $P$.
In other words, if morphing does not start at time $S^{e}_{p}(C^{e/p}_{p})$, edge $e$ can avoid crossing with an opposite edge $e/p$ at the intersection point $p$.
Function $P_{crit}^{(1)}:E\rightarrow 2^{U}$ provides the union of these periods for all intersection points on $e$.
If the morphing start time of edge $e$ is undefined, let $C^{e}_{p}=\emptyset$ and $S^{e}_{p}(\emptyset)=\emptyset$.
Therefore, $I_{s}(e)$ appearing on the right side of Eq.~(\ref{ep:Pcrit1}) may be replaced by $I(e)$.

Function $t_{earliest}:2^{U}\rightarrow U$ yields the minimum non-negative value not included in the time period $P$.
This definition can be expressed as in Eq.~(\ref{ep:earliestSafeTime}).

\begin{equation}
  t_{earliest}(P) = \min([0,\infty)\cap P^{c}) \label{ep:earliestSafeTime}
\end{equation}

One of the outputs from Alg.~\ref{alg:scheduleBasic}, $t_{total}$, is the total morphing time, which denotes the cycle length when morphing is repeated.

\section{Overlapping a part of each cycle}
\label{ssec:reducing_cycle1}

\begin{figure}[htb]
  \centering
  \includegraphics[scale=1.0]{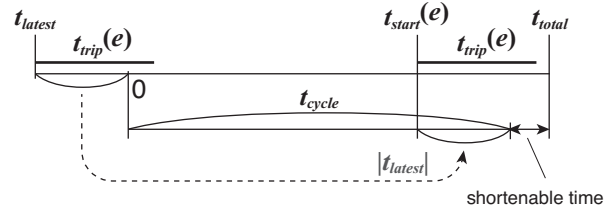}
  \caption{Schematic of concept for shortening a single cycle}
  \label{fig:cycle_time}
\end{figure}

Based on the schedule (that has already been determined), let $t_{latest}$ ($<0$) be the time before time zero when the morphing of edge $e$ can be started. Shortening the cycle by $t_{total} - (t_{start}(e) + |t_{latest}|)$ will not cause the crossing of $e$ (see Fig.~\ref{fig:cycle_time}).
In other words, the period can be shortened to $t_{start}(e) - t_{latest}$ without causing any crossings at edge $e$.
Overall, the graph can shorten the cycle to the maximum value at all edges $e\in E$.
Eq.~(\ref{ep:recudingCycle}) shows a shortened cycle length $t_{cycle}$.

\begin{equation}
t_{cycle} = \max_{e\in E} \{t_{start}(e) - t_{latest}(P_{fbd}^{(2)}(e))\}, \label{ep:recudingCycle}
\end{equation}
where $t_{latest}(P_{fbd}^{(2)}(e))$ provides the latest possible morphing start time before time zero of edge $e$.
Function $P_{fbd}^{(2)}$ is an extension of $P_{fbd}^{(1)}$, and its definition is given in Eqs.~(\ref{ep:Pfbd2}) and (\ref{ep:Pself1}).

\begin{align}
P_{fbd}^{(2)}(e) &= P_{crit}^{(1)}(e)\cup P_{self}^{(1)}(e) \label{ep:Pfbd2} \\
P_{self}^{(1)}(e) &= [t_{start}(e) - \tau_{trip}(e), t_{start}(e) + \tau_{trip}(e)) \label{ep:Pself1}
\end{align}
Function $P_{fbd}^{(2)}$ uses $P_{self}^{(1)}:E\rightarrow 2^{U}$ in addition to $P_{crit}^{(1)}$.
Function $P_{self}^{(1)}(e)$ yields the time period when morphing is prohibited to start so that it does not overlap with its own morphing.
When the start time $t_{start}(e)$ of edge $e$ is undefined, let $P_{self}^{(1)}(e)=\emptyset$.
Thus, function $P_{fbd}^{(2)}$ can be used instead of $P_{fbd}^{(1)}$.
Function $t_{latest}:2^{U}\rightarrow U$ yields the negative (or zero) upper bound that is not included in the time period $P$($\subseteq U$) given.
This function is defined in Eq.~(\ref{ep:nLatestTime}).

\begin{equation}
  t_{latest}(P) = \max ((-\infty,0) \cap P^{c}) \label{ep:nLatestTime}
\end{equation}

\section{Duplication within a Cycle}
\label{ssec:duplication}

Each stub stretches and shrinks only once within one cycle in the schedule obtained Alg.~\ref{alg:scheduleBasic} or Alg.~\ref{alg:scheduleBasic} plus Eq.~(\ref{ep:recudingCycle}).
However, some edges can be morphed two or more times within a single cycle, without causing any crossings.
In other words, focusing on certain edges may further reduce the average cycle length.

Hereafter, the start time of morphing with respect to an edge is treated as a set and is denoted by $T_{start}(e)$.
Accordingly, the definition of $I_{s}(e)$ is changed to $I_{s}(e)= \{p\in I(e)|T_{start}(e/p)\neq \emptyset\}$.
Function $P_{self}(e)$ is also extended.

Alg.~\ref{alg:scheduleDuplication} presents an algorithm for scheduling multiple morphs within one cycle.
The function $P_{fbd}^{(3)}:E\times \mathbb{R}_{\geq 0}\rightarrow 2^{U}$, defined by Eq.~(\ref{ep:Pfbd3}) yields the forbidden morphing start period of edge $e\in E$ when the cycle length is $c\in \mathbb{R}_{\geq 0}$.

\begin{algorithm}[htb]
  \caption{Scheduling multiple morphing within a single cycle}
  \label{alg:scheduleDuplication}
  \begin{algorithmic}[1]
    \Input{$E$ -- Set of edges in the morphing group, the start time $t_{start}(e)$ of all $e\in E$, $t_{total}$ -- total morphing time, $t_{cycle}$ -- morphing cycle length}
    \Output{Set of start times $T_{start}(e)$ for all $e\in E$}
    \Function{scheduleDuplication}{$E,t_{start},t_{total},t_{cycle}$}
      \For{$e$ in $E$}
        \State $T_{start}(e) \leftarrow \{ t_{start}(e) \}$
      \EndFor
      \State $E_{1} \leftarrow E$
      \While{$E_{1}\neq \emptyset$}
        \State $E_{2} \leftarrow \emptyset$
        \For{$e$ in $sort(E_{1})$}
          \State $t_{start2} \leftarrow t_{earliest}(P_{fbd}(e, t_{cycle}))$
          \If{$t_{start2} + \tau_{trip}(e)\leq t_{total}$}
            \State $T_{start}(e) \leftarrow T_{start}(e) \cup \{t_{start2}\}$
            \State $E_{2} \leftarrow E_{2} \cup \{e\}$
          \EndIf
        \EndFor
        \State $E_{1} \leftarrow E_{2}$
      \EndWhile
    \EndFunction
  \end{algorithmic}
\end{algorithm}

\begin{align}
P_{fbd}^{(3)}(e,c) &= W(c,P_{crit}^{(1)}(e)\cup P_{self}^{(2)}(e)) \label{ep:Pfbd3} \\
P_{self}^{(2)}(e) &= \bigcup_{t\in T_{start}(e)}[t - \tau_{trip}(e), t + \tau_{trip}(e)),
\end{align}
where the function $W:\mathbb{R}_{\geq 0}\times 2^{U} \rightarrow 2^{U}$ is defined as $W(c,P) = P\cup (\cup_{[a,b)\subseteq P}[a+c, b+c))$.
This adds one cycle length $c$ to the time period $P$.
If $c$ is equal to the $t_{total}$ obtained from Alg.~\ref{alg:scheduleBasic}, the function $W$ does not need to be applied.
However, if $c$ is shorter than $t_{total}$, there may be edges that morph across two cycles; therefore, the period is extended to two cycles to determine the forbidden morphing start period.

\section{Allowance of Crossings}
\label{ssec:allowing_crossings}

We considered scheduling that allows for up to a certain number of crossings ({\em allowable crossing number}) $n$ per edge.
Thus far, we proceeded with the explanation assuming that there were no always-passing intersections. However, hereafter, we include always-passing intersections in our considerations.

Let the {\em controllable crossing number} be the allowable crossing number minus the number of always-crossing intersections.
When the number of always-crossing intersections exceeded the allowable crossing number, let the controllable crossing number be zero.
Because we cannot control the occurrence of crossings at the always-crossing intersections, we perform scheduling by ignoring these crossings based on the controllable crossing number.

As crossings at semi-avoidable intersections cannot be avoided, the number of crossings may exceed the controllable crossing number. 
Even in these cases, the number of crossings should be maintained as low as possible.
For example, let us suppose two semi-avoidable intersections exist on an edge.
Although crossings at these intersections cannot be avoided, it may be possible to schedule them such that no two crossings occur simultaneously.

Function $P_{fbd}^{(4)}:E\times \mathbb{R}_{\geq 0}\times \mathbb{N}\rightarrow 2^{U}$, which determines the forbidden morphing start period of edge $e$ when the presence of always-passing intersections is allowed and when a certain number of crossings is allowed, can be expressed as in Eq.~(\ref{ep:Pfbd5}).

\begin{equation}
P_{fbd}^{(4)}(e,c,n) = W(c,P_{crit}^{(2)}(e,k_{n}(e)) \cup P_{self}^{(2)}(e) \cup P_{opst}(e,n)), \label{ep:Pfbd5}
\end{equation}
where $c\in \mathbb{R}_{\geq 0}$ represents the cycle length and is set to zero if undetermined.
The role of $W$ is the same as that described in \ref{ssec:duplication}.
The allowable crossing number $n\in \mathbb{N}$ given in advance is a common condition for all the edges. However, the controllable crossing number differs from edge to edge because the number of always-crossing intersections differs accordingly.
Therefore, let $k_{n}(e)$ denote the controllable crossing number for edge $e$. If a stub of edge $e$ always passes through intersection $p$, let $C^{e}_{p}=U$, even if the morphing schedule of edge $e$ remains undefined.
$P_{opst}:E\times \mathbb{N}\rightarrow 2^{U}$ is a function used to find the critical time period of conditions for opposite edges.
When no crossing is allowed, there is no need to consider these conditions, because when no crossing occurs for the target edge, the same condition applies to opposite edges as well.
However, when allowing crossings and setting their upper limits, conditions for the target edge differ from those for the opposite edges.
We explain the definition of the $P_{opst}$ function in \ref{ssec:opst}.

\subsection{Satisfying Allowable Crossing Number for Target Edge}

The critical period during which the controllable crossing number of edge $e$ exceeds $k$ is represented by $P_{crit}^{(2)}(e,k)$, as indicated in Eq.~(\ref{eq:Pcrit3}).

\begin{align}
P_{crit}^{(2)}(e,k) &=
  \begin{cases}
    P_{cirt}^{(1)}(e)\cup\{\bigcup_{Q\in I_{sa}(e)^{\#2}} P_{cSub}(e,Q)\} & \text{if $k=0$} \\
    \bigcup_{Q\in I_{s'}(e)^{\#k+1}} P_{cSub}(e,Q) & \text{otherwise}
  \end{cases} \label{eq:Pcrit3} \\
P_{cSub}(e,Q) &= 
  \begin{cases}
    \emptyset & \text{if $O(e,Q)=U$} \\
    \bigcap_{p\in Q} S^{e}_{p}(O(e,Q)) & \text{otherwise}
  \end{cases} \label{eq:Pesub} \\
O(e,Q) &= \bigcap_{p\in Q} C^{e/p}_{p}, \label{eq:overlapped}
\end{align}
where $I_{a}(e) = \{p\in I(e)|p\mbox{ is avoidable}\}$, 
$I_{sa}(e) = \{p\in I(e)|p\mbox{ is semi-avoidable}\}$, and
$I_{s'}(e)= I_{s}(e)\cup \{p\in I_{a}(e)|e/p \mbox{ always passes } p\}$.
We set $Q$ in Eqs.~(\ref{eq:Pcrit3}) and (\ref{eq:Pesub}) as the subset of intersections on $e$ with two or $k+1$ elements.
$O(e, Q)$ denotes the period during which the stubs of the opposite edges pass simultaneously at $\#(Q)$ points on edge $e$.
If $O(e, Q)\neq \emptyset$, then crossings may occur simultaneously at all the intersection points in $Q$.
To avoid this, $e$ should not pass through these points during the time period.
$P_{cSub}(e,Q)$, shown in Eq.~(\ref{eq:Pesub}), represents the critical time period when a stub of edge $e$ starts and then passes through all the intersection points in that time period.
However, in Eq.~(\ref{eq:Pesub}), $P_{cSub}(e,Q)=\emptyset$ when $O(e, Q)=U$.
When the existence of always-passing intersections is allowed, unavoidable crossings may occur.
Case $O(e,Q)=U$ represents an unavoidable situation.
This indicates that all the opposite edges at the $\#(Q)$ intersection points in $Q$ always pass and that simultaneous crossings with all of them are unavoidable.
Therefore, for crossings at intersection points $Q$, the critical time period is $\emptyset$, and it does not affect the start time.

\subsection{Satisfying Allowable Crossing Number for Opposite Edges}
\label{ssec:opst}

The critical time period in which the number of crossings of the opposite edges of edge $e$ exceeds the allowable crossing number $n$ is represented by $P_{opst}(e,n)$, as indicated by Eqs.~(\ref{eq:p_other2}) and (\ref{eq:PoSub2}), where $I_{a1}(e) = \{p\in I(e)|p\mbox{ is avoidable for }e\}$.

\begin{align}
P_{opst}(e,n) &= \bigcup_{p\in I_{a1}(e)} S_{p}^{e} ( P_{oSub}(e/p,p,n) ) \label{eq:p_other2} \\
P_{oSub}(e',p,n) &= 
  \begin{cases}
    C^{e'}_{p} \hspace{22mm} \text{if $k_{n}(e')=0 \land$ ($p$ is avoidable for $e'$)} \\
    \bigcup_{q\in I_{a}(e')} X(q) \hspace{5.3mm} \text{if $k_{n}(e')=0 \land$ ($p$ is always-passing for $e'$)} \\
    \bigcup_{Q\in I_{s'}(e')^{\#k_{n}(e')}} \bigcap_{q\in Q} X(q) \hspace{10mm} \text{otherwise,}
  \end{cases} \label{eq:PoSub2}
\end{align}
where $X(q)$ denote the time period when a crossing occurs at point $q$.
That is, $X(q) = C_{q}^{e_{1}}\cap C_{q}^{e_{2}}$ when $e_{1}$ and $e_{2}=e_{1}/q$ cross at point $q$.
$P_{oSub}(e', p, n)$ represents the critical time period at intersection $p$ with respect to an opposite edge $e'$ for the allowable crossing number $n$, as shown in Eq.~(\ref{eq:PoSub2}).
The right-hand side of Eq.~(\ref{eq:p_other2}) indicates the union of the forbidden morphing start periods when $e$ passes through the intersection point with the opposite edge during this critical time period.
The definition of $P_{oSub}$ can be divided into three cases.
(1) If $k_{n}(e')=0$ and $e'$ can avoid $p$, then the period $C^{e'}_{p}$ (at which $e'$ passes intersection point $p$) is the critical time period.
(2) If $k_{n}(e')=0$ and $e'$ always passes through $p$, then $e$ should be allowed to pass through $p$, provided that all crossing time periods at semi-avoidable intersections that are avoidable for $e'$ are avoided.
This implies that the critical time period is the time period of crossing occurrence at the avoidable intersections for $e'$.
(3) Otherwise, the critical time period is the time period in which more than $k_{n}(e')$ crossings occur simultaneously on $e'$.

\subsection{Overlapping a part of each cycle}
\label{ssec:reducing_cycle2}

The method proposed in \ref{ssec:reducing_cycle1} shortens the cycle length by determining the possible start time of the following cycle for each edge.
The possible start time of each single edge was examined, however, the effect of shifting the start time of all the edges was not inspected.
Therefore, if the allowable crossing number is greater than or equal to one, the method does not function properly.

Because we have not yet identified an efficient method to address this issue, we only present a simple countermeasure.
The method involves affording a tentative shortened cycle length using the method described in \ref{ssec:reducing_cycle1}, then checking the time period when the condition is violated, and extending the cycle length by the amount of time when the condition is violated.

\section{Evaluation of Effectiveness}

We implemented the scheduling algorithm described in Alg.~\ref{alg:scheduleBasic} and Alg.~\ref{alg:scheduleDuplication} along with the functions in Java with JRE 16.0.2.
The cycle length was defined as a real number (an element of $\mathbb{R}_{\geq 0}$) in the aforementioned explanations; however, in our implementation, it was defined as an \texttt{int} with ms as the unit.

Experiments were conducted to investigate the effects of each of the previously described factors: overlapping a part of each cycle, duplicating morphs in one cycle, and allowance of crossings.
We prepared complete graphs with 7--13 nodes and laid out the nodes of each graph equally spaced around a circumference with a radius of 200 pixels.
The speed of the tips of the stubs was 100 pixels/s and the tips were paused for 100 ms at the longest stub ratio.

The longest stub-ratio was set to $\eta=50\%$ and the shortest stub-ratios $\delta$ were set to 4\%, 9\%, 16\%, and 25\%.
Always-passing intersections were included in the case of $\delta=25\%$ with seven nodes, $\delta\geq 16\%$ with 8--10 nodes, and $\delta\geq 9\%$ with 11--13 nodes.
Furthermore, always-crossing intersections were included in the case of $\delta=25\%$ with 10--13 nodes.

For each of these 28 combinations, morphing scheduling was performed with or without the application of overlapping, duplication within a cycle, and by changing the allowable crossing number from 0 to 10.
In addition, considering the effects of $sort$ in Alg.~\ref{alg:scheduleBasic}, scheduling was performed with 100 different orders under the same conditions, including a descending order of the edge lengths and vice versa, as well as 98 randomly sorted orders.

\subsection{Overlapping a part of each cycle}

\begin{figure}[htb]
\begin{minipage}[t]{0.45\hsize}
\centering
\includegraphics[scale=0.60]{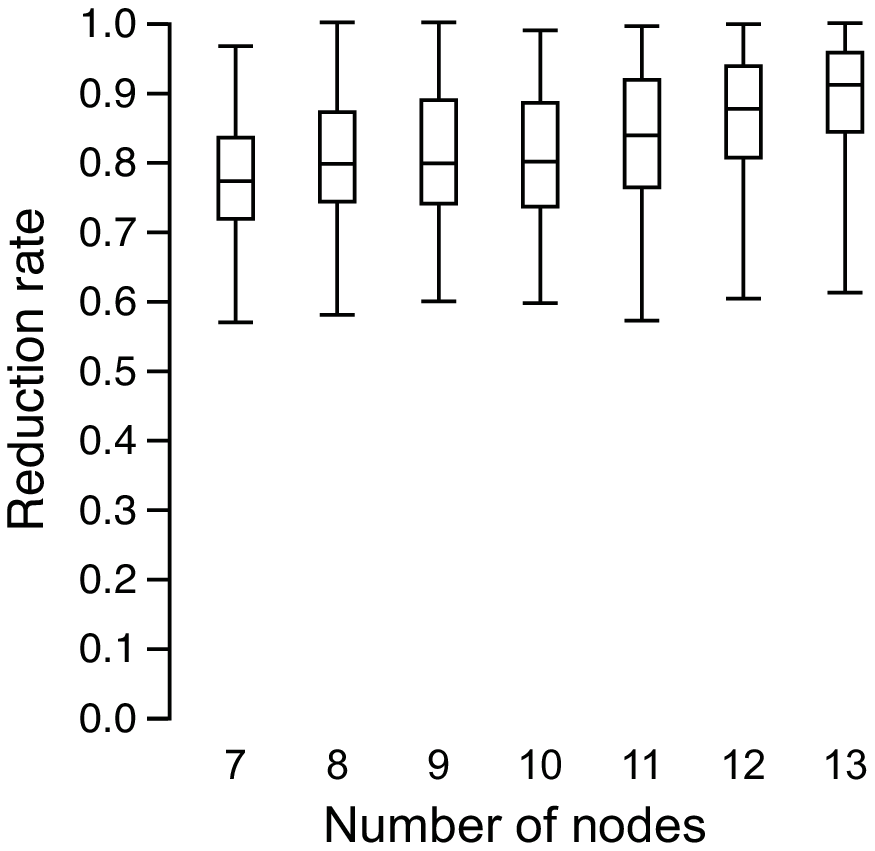} \\
(a) Overlapping a part of each cycle
\end{minipage}
\hspace{0.05\hsize}
\begin{minipage}[t]{0.45\hsize}
\centering
\includegraphics[scale=0.60]{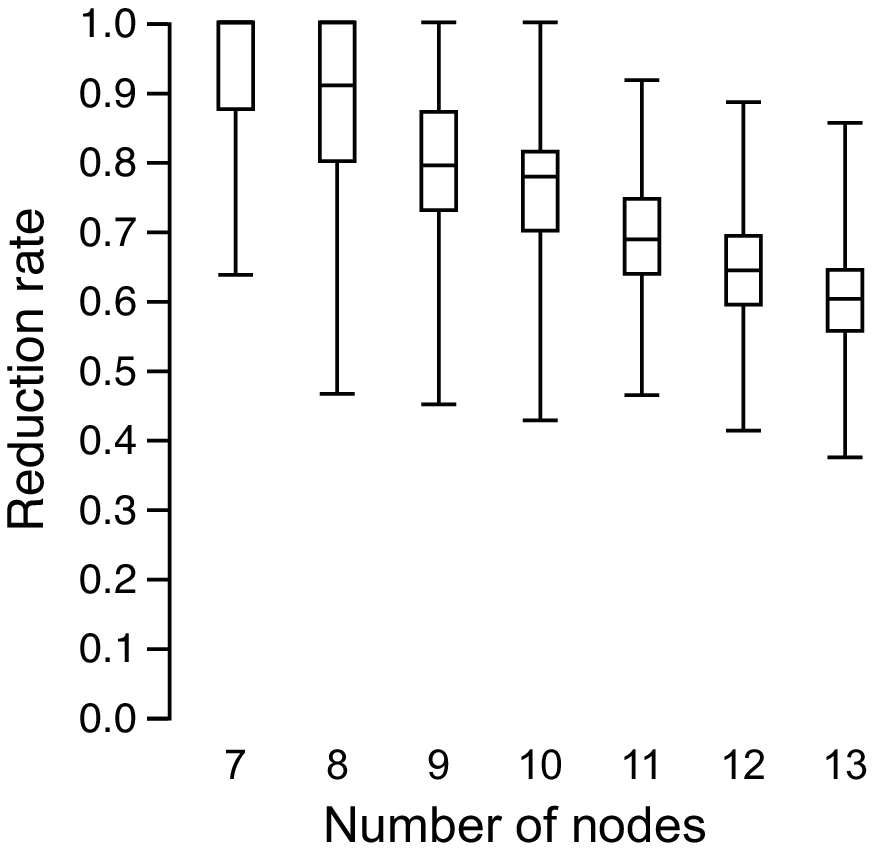} \\
(b) Duplication within a cycle
\end{minipage}
\caption{Effects of overlapping a part of each cycle and duplication within a cycle}
\label{fig:boxp-shortening-n-dup}
\end{figure}

Here, we examined the reduction rate of the cycle length and derived it as the ratio of the cycle length obtained by applying the proposed methods to the cycle length obtained by the scheduling algorithm proposed by Misue \& Akasaka \cite{misue2019-graph}.
The number of samples was 2,400 for seven nodes, 3,600 for eight nodes, and 4,400 for each of the other cases.
Fig.~\ref{fig:boxp-shortening-n-dup}(a) shows the quartiles of the reduction rates of cycle lengths.
Although a certain effect is observed, it is found that this decreases as the number of nodes increases.
The median value for the seven nodes is 0.771, but it increased to 0.910 for 13 nodes.

\subsection{Duplication within a Cycle}

For example, if all the edges could morph twice within one cycle, the cycle length would be effectively halved.
Hence, we considered the reduction rate as the number of edges to be morphed divided by the total number of morphs.
The number of samples was the same as that used for the evaluation of the overlapping cycles.
In all the cases, the overlapping a part of each cycle was not applied.

Fig.~\ref{fig:boxp-shortening-n-dup}(b) shows the quartiles of the reduction rates. 
It can be observed that the effectiveness improves as the number of nodes increases.
Focusing on the median, when the number of nodes is seven, the median is one, and there is no reduction effect; however, when the number of nodes is 13, the median is 0.602. In other words, on an average, morphing can be performed nearly twice within one cycle.

\subsection{Allowance of Crossings}

\begin{figure}[htb]
\begin{minipage}[t]{0.45\hsize}
\centering
\includegraphics[scale=0.60]{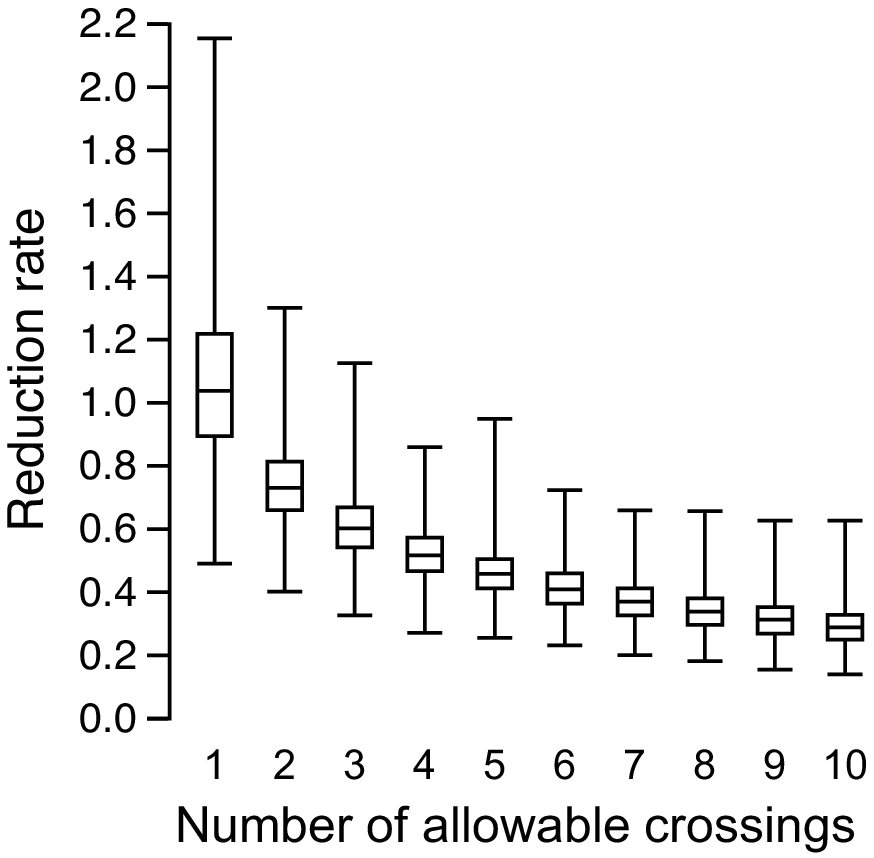} \\
(a) All types of intersections
\end{minipage}
\hspace{0.05\hsize}
\begin{minipage}[t]{0.45\hsize}
\centering
\includegraphics[scale=0.60]{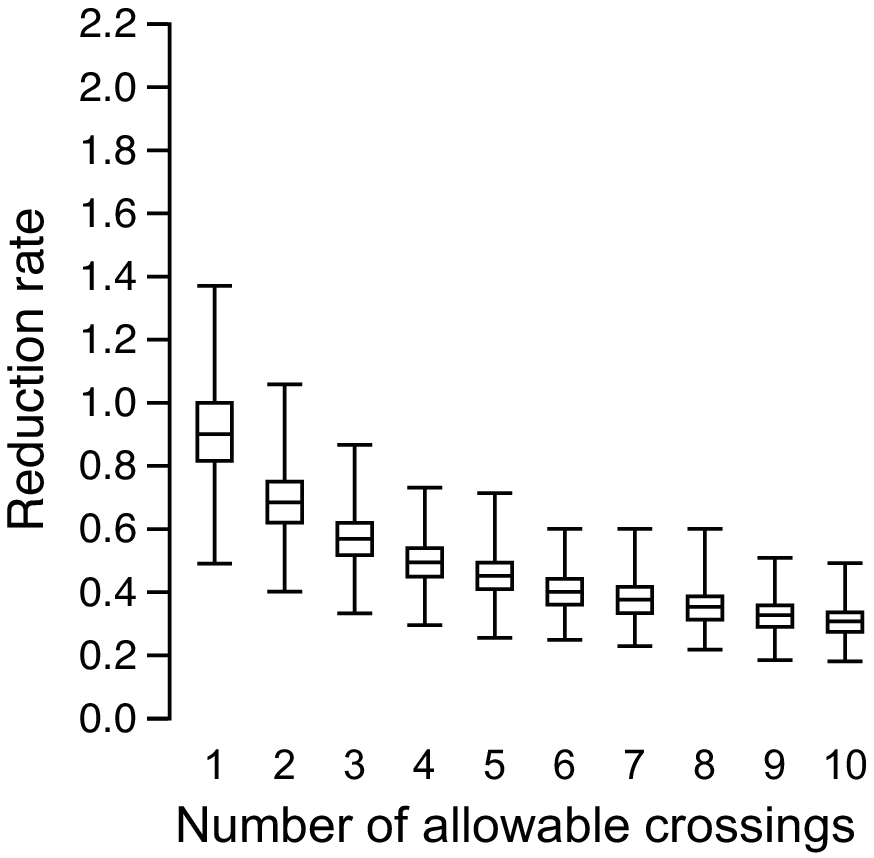} \\
(b) Fully avoidable intersections only
\end{minipage}
\caption{Effects of allowance of crossings}
\label{fig:boxp-nallowable2}
\end{figure}

We examined the reduction rate of the cycle length when crossings were allowed for each allowable crossing number.
Fig.~\ref{fig:boxp-nallowable2} shows the quartiles of the reduction rate of cycle lengths.
Fig.~\ref{fig:boxp-nallowable2}(a) shows the case in which all types of intersections are included, and Fig.~\ref{fig:boxp-nallowable2}(b) shows the case in which only fully avoidable intersections are included.
In both cases, the reduction effect improved as the allowable crossing number increased.
However, in some cases, the cycle becomes longer around the allowable crossing numbers of 1 to 3.
In the case of fully avoidable intersections, these cases are less frequent. 
In any case, the identification of the factors will be our focus in future studies.

\section{Conclusion}

We developed three scheduling methods to shorten the morphing cycle in MED.
The first method shortens a cycle by overlapping the end of the current cycle with the succeeding one.
The second method shortens the average duration of a cycle by duplicating every morph by the allowable number of times in one cycle.
The third method aims at shortening the cycle length by allowing a certain number of crossings at each edge.
We incorporated these developed methods into a program and conducted evaluation experiments on complete graphs laid out on a circle to confirm the effectiveness of each method. 

\subsubsection{Acknowledgements}
This work was supported by JSPS KAKENHI Grant Number JP21K11975.

%
%
%
 \bibliographystyle{splncs04}
 \bibliography{reference}



\end{document}